\documentclass[12pt]{article}

\usepackage[T1]{fontenc}
\usepackage{amsmath, bm}
\usepackage{amssymb}
\usepackage{tabularx}
\usepackage[dvips]{epsfig}
\usepackage[dvips]{graphics}
\usepackage{dsfont}
\usepackage{url}
\usepackage{slashbox}
\usepackage{natbib}
\usepackage{multirow}
\usepackage[dvipsnames]{xcolor}
\usepackage[normalem]{ulem}
\usepackage{float}
\usepackage{booktabs,subcaption,amsfonts,dcolumn}

\makeatletter
\newcommand*{\rom}[1]{\expandafter\@slowromancap\romannumeral #1@}
\makeatother

\usepackage{epstopdf} 
\usepackage{color} 
\usepackage[normalem]{ulem} 

    \newcommand{\tf}[1]{#1} 
    \newcommand{\tr}[1]{#1} 
   \newcommand{\trr}[1]{#1} 
    \newcommand{\ta}[1]{#1} 
    \newcommand{\tO}[1]{#1} 
    \newcommand{\td}[1]{} 

\newtheorem{theorem}{Theorem}[section]

\DeclareMathAlphabet{\mathcal}{OMS}{cmsy}{m}{n}
\SetMathAlphabet{\mathcal}{bold}{OMS}{cmsy}{b}{n}

\newcommand{\vertiii}[1]{{\left\vert\kern-0.25ex\left\vert #1 
    \right\vert\kern-0.25ex\right\vert_F}}

\title{High-Dimensional Functional Mixed-effect Model for Bilevel Repeated Measurements}
\author{Xiaotian Dai $^1$ \and Guifang Fu$^{1}\footnote{Corresponding Author: gfu@binghamton.edu}$}
\date
{%
    $^1$Department of Mathematical Sciences, SUNY Binghamton University, Vestal, NY 13850\\%
}

\begin{document}

\def\spacingset#1{\renewcommand{\baselinestretch}%
{#1}\small\normalsize} \spacingset{2}

\begin{tiny}
\emph{ }\\

Vol. xx, No. yy, 20??, Pages 1?25

\emph{}
\end{tiny}


{\let\newpage\relax\maketitle}

\begin{abstract}
\tf{The bilevel functional data under consideration has two sources of repeated measurements. One is to densely and repeatedly measure a variable from each subject at a series of regular time/spatial points, which is named as functional data. The other is to repeatedly collect one functional data at each of the multiple visits. Compared to the well-established single-level functional data analysis approaches, those that are related to high-dimensional bilevel functional data are limited.} In this article, we propose \tr{a high-dimensional functional mixed-effect model (HDFMM)} to \tf{analyze the association between the bilevel functional response and a large scale of scalar predictors.} \tf{We utilize B-splines to smooth and estimate the infinite-dimensional functional coefficient, a sandwich smoother to estimate the covariance function, and integrate the estimation of covariance related parameters together with all regression parameters into one framework through a fast updating MCMC procedure.} We demonstrate that the performance of the HDFMM method is promising under various simulation studies and a real data analysis.  As an extension of the well-established linear mixed model, the \tr{HDFMM} model extends the response from repeatedly measured scalars to repeatedly measured functional data/curves, while maintaining the ability to account for relatedness among samples and control for confounding factors. 
\end{abstract}

\noindent%


\section{INTRODUCTION}
\tf{  \ta{The importance of bilevel functional data, where multiple \tO{correlated} functional data are repeatedly measured on the same subject at multiple visits, has already been recognized \citep{di2009multilevel, chen2012modeling, morris2003wavelet, crainiceanu2009}.}  For example, \ta{electroencephalographic (EEG) curves were densely measured during a subject's sleeping hours and collected twice for each subject at two different years, which is the well-known sleep heart health study (SHHS)} \citep{di2009multilevel}. Our motivating dataset consists of twenty-five leaves repeatedly collected from each \emph{Populus euphratica} tree from various directions and heights above ground. Each leaf shape was described as a 910-dimensional functional data. 

The scientific question is to detect important predictors influencing the variation of dynamic trajectories of the curves (as the response), or in predicting the curve trajectories for new observations. Some common challenges for today's bilevel functional data analysis include:} \tf{First, high dimensionality comes from multiple different sources, such as the number of time/spatial points of each functional data, the number of repeatedly collected functional data for each subject, and the number of predictors can be larger than the number of subjects ($p>n$).} \tf{Second, correlations exist in different time/spatial points of each functional data, different functional data repeatedly measured for each subject, and different predictors of the same subject.} Third, not only between-subject but also within-subject heterogeneity and variation exist in bilevel functional data.

To overcome these challenges, we propose \tr{a high-dimensional functional mixed-effect model  (HDFMM)} to quantify the association between \tf{the bilevel functional data response curves and a set of high-dimensional predictors}. The \tr{HDFMM} model \tf{has far-reaching application flexibility and} it produces multiple outputs: 1) The estimates and corresponding confidence intervals of all unknown parameters contained in the model. In particular, the estimate of a functional coefficient demonstrates how a predictor influences the dynamic trajectory of bilevel functional data response varying over time. 2) It encourages sparsity and shrinks the coefficients of unimportant predictors into zero. 3) Several pieces of curve-related information: \tf{estimated overall mean curve across all observations, predicted dynamic trend for every repeatedly measured functional data at each visit for each subject, and their variation modes.} The \tf{performance} of the proposed \tr{HDFMM} approach is tested via multiple simulations and one real data analysis.

\tf{There exists a large amount of well-established functional mixed effect models} \citep{james2002generalized, muller2005generalized, yao2005functional, ramsay2006, crainiceanu2009, zhu2012, li2015}. See \cite{morris2015functional}, \cite{greven2017general}, and \cite{morris2017comparison} for comprehensive reviews for more ``function-on-scalar'' regression models. However, approaches related to \tf{high dimensional bilevel functional data} are still limited. \tf{The proposed HDFMM model provides an alternative option to current functional mixed effect literature, but it differs from other relevant approaches in the following aspects: The Bayesian group Lasso approach proposed by \cite{li2015} also detected the association of functional curves with a set of high-dimensional predictors. However, they focused on independent curves without having the mixed structure driven by repeated measurements. The multilevel functions considered in the SHHS data \citep{di2009multilevel, crainiceanu2009} and the mortality data \citep{chen2012modeling} actually have two sets of time: the time measurement for each curve, and the time over the repeated collections of multiple curves. Compared to their setting of `longitudinally-collected multilevel functional data', the HDFMM focuses on repeatedly-collected bilevel functional data that has only one set of time/spatial measurement. We adopt the ideas of \cite{di2009multilevel} and \cite{morris2003wavelet} for hierarchical functional modeling (without involving any predictors), but HDFMM focuses more on the association of curve response and predictors. In terms of the other multilevel function-on-scalar mixed effect models \citep{guo2002functional, zhu2012, goldsmith2015generalized}, the HDFMM model has made improvements to focus on high dimensionality as well as variable selection purposes. Many other approaches also regularized the functional coefficients of predictors to cope with high dimensionality \citep{morris2006wavelet, scheipl2015functional, scheipl2016identifiability}; however, they modeled the random effect terms by a multivariate Gaussian approach, which is different from the nonparametric random process in the HDFMM.}

\tf{The remainder sections are organized as follows: In Section 2, we detail our proposed HDFMM model, including its theoretical Bayesian properties. This is followed by an assessment in Section 3 of the finite sample performance of HDFMM via simulation studies and a real data analysis. In Section 4, we discuss the method and give our conclusion. Finally, the proof of the main theorem is given in the appendix.}

\section{MODELS AND INFERENCE PROCEDURES}
Let $\{Y_{ij}(t_{k}), i=1,\ldots,I;~j=1,\ldots,J;~k=1,\ldots,$ $K;~t_{k}\in [0,1]\}$ denote the bilevel functional data curve response measured at \tf{a sequence of time/spatial points} $t_{k}$ for the $j^{th}$ repeated \tr{measurement collected from} the $i^{th}$ subject, where $I$ is the number of subjects, $J$ is the number of repeated measurements for each subject, and $K$ is the number of grids of each curve. Let $p$ be the number of predictors, and $q$ be the number of covariates.

\tr{The \tr{HDFMM} model is constructed as
\begin{equation}
Y_{ij}(t) =  \boldsymbol{\beta}^T(t) \boldsymbol{X_i} + \boldsymbol{C}^T(t) \boldsymbol{W_{i}} +U_{ij}(t) + \epsilon_{ij}(t),
\label{main}
\end{equation}
where $\boldsymbol{X_i} = [X_{i1}, \ldots, X_{ip}]^T$ is the high-dimensional predictor vector for subject $i$, and $\{X_{ir}^c,~ r=1,\ldots,q-1\}$ are covariates that are included when there is a need for control purposes, such as age, treatment, or population structure. Define $\boldsymbol{W_{i}}=[1, X_{i1}^c, \ldots, X_{i,q-1}^c]^T$} as the vector that does not need to be regularized, including the intercept and covariates. Here $\boldsymbol{\beta}(t) = [\beta_1(t), \ldots, \beta_p(t)]$ is the $p$-dimensional functional coefficient specifying the effect of predictors $\boldsymbol{X_i}$ on the response curve; and $\boldsymbol{C}(t)$ is the $q$-dimensional functional coefficient for $\boldsymbol{W_{i}}$. $U_{ij}(t)$ is the residual subject- and visit-specific deviation from the subject-specific mean, and $\epsilon_{ij}(t)$ is an i.i.d. white noise process with a constant variance $\sigma_{\epsilon}^2$ \citep{di2009multilevel, crainiceanu2009}. In (\ref{main}), the subject-specific effects $\boldsymbol{\beta}^T(t)$ and $\boldsymbol{C}^T(t)$ are treated as fixed functions, while the variance component $U_{ij}(t)$ is \tf{random} as mean 0 stochastic processes. We assume that $U_{ij}(t)$ and $\epsilon_{ij}(t)$ are uncorrelated.
The predictors and covariates have separate design matrices because \tf{it} enables users to flexibly choose which independent variables to regularize.

\subsection{Parameterization}
\subsubsection{Parameterization of functional coefficients}
\ta{To \tO{model} the functional coefficient curve $\boldsymbol{\beta}(t)$ contained} in (\ref{main}), we apply the non-parametric basis spline (i.e., B-spline) method to approximate the infinite-dimensional \tf{functional coefficient}. \ta{The functional coefficients of the $m^{th}$ predictor and $r^{th}$ covariate are modeled as
\tr{\begin{equation}
\begin{split}
\beta_m(t) & = \boldsymbol{b_m}^T \boldsymbol{\Phi}(t), \quad m = 1,...,p, \\
C_r(t) & = \boldsymbol{c_r}^T\boldsymbol{\Phi}(t), \quad r = 1,..., q,
\end{split}
\label{bk}
\end{equation}}
where \tr{$\boldsymbol{\Phi}(t) = [\Phi_1(t), ..., \Phi_v(t)]^T$} are $v$-dimensional basis functions of $t \in [0,1]$, and $\boldsymbol{b_m} = (b_{m1}, ..., b_{mv})$ and $\boldsymbol{c_r} = (c_{r1}, ..., c_{rv})$ are $v$-dimensional row vectors of corresponding expansion coefficients, respectively.

}

\subsubsection{Parameterization of covariance and eigenfunctions}

\ta{As a mean 0 stochastic process}, $U_{ij}(t)$ can be decomposed by the Karhunen-Lo\`eve (KL) expansion as \citep{di2009multilevel, crainiceanu2009}
\[
U_{ij}(t) = \sum_{l = 1}^{\infty} \zeta_{ijl}\phi_l(t),
\]
where $\{\phi_l(t)\} = \{\phi_1(t), \phi_2(t), \ldots \}$ are eigenfunctions \ta{forming} an orthonormal basis of $L_2[0, 1]$, $\{\zeta_{ijl}\} = \{\zeta_{ij1}, \zeta_{ij2}, \ldots \}$ are principal component (PC) scores with $E(\zeta_{ijl}) = 0$, and $Var(\zeta_{ijl}) = \lambda_l$, and $\{\lambda_l, l=1,\ldots \}$ are corresponding ordered eigenvalues with $\lambda_1 \geq \lambda_2 \geq \cdots$. For \ta{approximation and estimation purposes}, the infinite dimensionality of KL expansion is truncated to a finite number $L$ based on the proportion of \ta{total variation} explained by the first $L$ principal components. \ta{Then (\ref{main}) can be re-written as 
\tr{
\begin{equation}
\begin{cases}
Y_{ij}(t) = [\boldsymbol{\Phi}^T(t) \boldsymbol{b_1}, \ldots, \boldsymbol{\Phi}^T(t) \boldsymbol{b_p}] \boldsymbol{X_i} + [\boldsymbol{\Phi}^T(t) \boldsymbol{c_1}, \ldots, \boldsymbol{\Phi}^T(t) \boldsymbol{c_{q}}] \boldsymbol{W_{i}} +  \\\quad\quad+ \sum_{l = 1}^{L} \zeta_{ijl}\phi_l(t) + \epsilon_{ij}(t), \\
  \zeta_{ijl} \sim N(0, \lambda_l); \quad \epsilon_{ij}(t) \sim N(0, \sigma_{\epsilon}^2), \\\quad\quad i = 1, \ldots, I; ~j = 1, \ldots, J; l=1,\ldots,L; ~t \in [0,1].
\end{cases}
\label{Udecomp}
\end{equation}}}
The variable $\{\zeta_{ijl}\}$ and unknown parameter $\{\lambda_l\}$ contained in  (\ref{Udecomp}) \ta{will both be modeled as random variables and estimated jointly with all other unknown parameters via a Bayesian framework}. Since the real data in practice are contaminated with measurement errors, directly using the eigenvectors of the empirical sample covariance matrix $\Sigma_U$ tends to be noisy, which affects the accuracy of the results. Therefore, we estimate a smooth version (denoted as $\hat{G}(t_{k_1},t_{k_2})$) of the empirical covariance matrix $\Sigma_U$, applying the sandwich smoother approach proposed by \cite{xiao} and the fast covariance estimation (FACE) algorithm described in \cite{xiao2013}. Situations where $K>500$ pose significant challenges to covariance matrix smoothing in terms of computational and storage burdens for high-dimensional matrices \citep{xiao2013}. To tackle this challenge, FACE implements \tf{a fast bivariate smoothing method}, speeds up the computation, and realizes the high-dimensional covariance function smoothing in functional data analyses \citep{xiao2013}. Once the covariance function $\hat{G}(t_{k_1},t_{k_2})$ is smoothed and estimated, its eigendecomposition provides the eigenfunction estimates, i.e., $\{\hat{\phi}_l(t);l=1,\ldots,L\}$ \citep{yao2005}. 

\subsubsection{Tuning parameters}
\ta{The first uncertainty of the \tr{HDFMM} method is related to the choice of $L$ in (\ref{Udecomp}). We applied a simple rule to choose $L$ as proposed by \cite{di2009multilevel}, which simultaneously satisfies two criteria}: 1) the proportion of \ta{total variation} explained by the first $L$ principal components is larger than a \ta{pre-determined} threshold (e.g., 90\%); 2) the proportion of variation explained by any additional principal component is less than another \ta{pre-determined} threshold (e.g., 1\%). This rule works well in simulation studies and in practical analyses \citep{di2009multilevel}. The second uncertainty in the model comes from the choice of a B-spline basis system, i.e., the number of basis functions (i.e., value of $v$), the number of knots, and the degree of B-spline functions. \cite{de1978} and \cite{ramsay2006} provided detailed descriptions regarding the number of knots to be selected and how to position them. After the knots are chosen, the degree of B-spline functions can be decided using model selection criterion such as BIC \citep{li2015}.

\subsection{Bayesian inference procedures}
Equation (\ref{Udecomp}) involves a large number of unknown parameters, and extra challenges arise when the number of predictors is much larger than the number of observations. Even though the candidate predictor pool is extremely large, only a few predictors are truly influential to the response, which is the common sparsity property \citep{zhou2013polygenic}. The predictors should therefore be regularized, and parameters of inactive predictors should be shrunk towards zero to prohibit overfitting and improve prediction accuracy. 

\tr{The HDFMM method does not regularize the functional coefficients of high-dimensional predictors directly through smoothing the theoretically infinite-dimensional functions as did in \cite{scheipl2016identifiability}. Instead, the group Lasso-based regularization technique is applied to the functional coefficients of the spline basis at the group level. The ``group''  here refers to \tf{the multiple components of regression coefficients of the spline basis for each predictor, and its theoretical advantages were claimed by} \citet{yuan2006model}.} As an extension of LASSO, \cite{yuan2006model} proposed a group Lasso approach that partitions parameters into multiple subgroups and imposes a constraint on the sum of the $L_2$ norm of each subgroup. In (\ref{Udecomp}), all of the $v$ components of each expansion coefficient vector, $\boldsymbol{b_m} = (b_{m1}, ..., b_{mv})^T$ naturally belong to the same subgroup because they are all related to the same predictor and should share some common properties. Let $||\boldsymbol{b_m}||,~m=1,\ldots,p,$ be the $L_2$ norm of each coefficient vector. The effect of the $m^{th}$ predictor is zero if and only if $||\boldsymbol{b_m}|| = 0$. We choose not to penalize the coefficients of $\boldsymbol{c_r}$'s to expand model's flexibility because the covariate is usually used for control or comparison purposes, and therefore has a small dimension. 
  
Based on the group Lasso scheme, the regularization process on the high dimensional coefficients, $\boldsymbol{b_m}$, is performed by minimizing:
\tr{
\begin{equation}
arg\,min_{\boldsymbol{b_m}} (\frac{1}{2} ||\boldsymbol{y} - \boldsymbol{\hat{y}}||^2 + \lambda_R \sum_{m = 1}^{p} ||\boldsymbol{b_m}||),
\end{equation}
where $\boldsymbol{y}$ is the $I \times J \times T$-dimensional vector reshaped from $Y_{ij}(t)$ in  (\ref{Udecomp}) across all values of $i = 1, \ldots, I; ~j = 1, \ldots, J;$ and $t \in [0,1]$.
And $\boldsymbol{\hat{y}}$ is the predicted version of $\boldsymbol{y}$, and $\lambda_R$ is a tuning parameter.} Bayesian group Lasso models used multivariate Laplace priors for the coefficients belonging to each subgroup \citep{li2015}. Multivariate Laplace priors have an advantage in that they can pull the coefficients of non-influential predictors to zero faster than a multivariate normal prior or a Student-$t$ prior \citep{park2008bayesian}. Therefore, we apply a $v$-dimensional multivariate Laplace prior for \ta{\tO{each expansion coefficient vector}, corresponding to one subgroup in the group Lasso model, as}:
\begin{center}
$\mbox{M-Laplace}(\boldsymbol{b_m} | \boldsymbol{0}, (v\lambda_R^2 / \sigma_{\epsilon}^2) ^ {-\frac{1}{2}}) = (v\lambda_R^2 / \sigma_{\epsilon}^2) ^ {\frac{v}{2}}\mbox{exp}(-(v\lambda_R^2 / \sigma_{\epsilon}^2) ^ {\frac{1}{2}} ||\boldsymbol{b_m}||).$
\end{center}

A multivariate Laplace prior can be rewritten as a scale mixture of multivariate normal and Gamma distributions:
\begin{equation}
\begin{split}
\mbox{M-Laplace}&(\boldsymbol{b_m} | \boldsymbol{0}, (v\lambda_R^2 / \sigma_{\epsilon}^2) ^ {-\frac{1}{2}}) \\
&\propto \int_{0}^{\infty} \mbox{MVN}(\boldsymbol{b_m}|\boldsymbol{0}, \sigma_{\epsilon}^2 \tau_m^2) \mbox{Gamma}(\tau_m^2|\frac{v+1}{2}, \frac{2}{v\lambda_R^2})d\tau_m^2,
\end{split}
\label{laplace}
\end{equation}
where \ta{$(v+1)/2$ and $2/(v\lambda_R^2)$ are the shape and scale parameters of a Gamma distribution, respectively; \ta{and $\tau_m^2$ is a latent transition \tO{hyperparameter}}. See \cite{raman2009} for the detailed derivations}. The mixture of the two distributions described in (\ref{laplace}) guarantees that the priors of all parameters under penalization can be expressed as the following hierarchical structure
\begin{center}
$\boldsymbol{b_m}|\sigma_{\epsilon}^2, \tau_m^2 \sim \mbox{MVN}(\boldsymbol{0}, \sigma_{\epsilon}^2 \tau_m^2),$ \quad \quad \text{and}  \quad $\tau_m^2|\lambda_R^2 \sim \mbox{Gamma}(\frac{v+1}{2}, \frac{2}{v\lambda_R^2}).$ \\
\end{center}
We also assume Gamma priors for the tuning parameter \citep{gelman2006prior}, 
\[
\lambda_R^2 \sim \mbox{Gamma}(\alpha_{1R}, \alpha_{2R}).
\]

In addition to the aforementioned high-dimensional parameter setting, we use standard Bayesian settings for all the other parameters that do not need to be penalized or regularized as follows
\begin{center}
$\boldsymbol{c_r} \sim \mbox{MVN}(\boldsymbol{0}, \Sigma_{\boldsymbol{c_r}}),$ \quad \quad $\zeta_{ijl} \sim \mbox{N}(0, \lambda_l), \quad \quad \text{and}  \quad \lambda_l \sim \mbox{IG}(\alpha_{1l}, \alpha_{2l}),$ \\
\end{center}
where $\Sigma_{\boldsymbol{c_r}}$ is the covariance matrix of $\boldsymbol{c_r}$, whose prior is set as a $v \times v$ identity matrix.  $\mbox{IG}(\alpha_{1l}, \alpha_{2l})$ is an inverse gamma distribution with shape parameter \tO{$\alpha_{1l}$} and scale parameter \tO{$\alpha_{2l}$}. We choose small values for \tO{$\alpha_{1l}$'s} and \tO{$\alpha_{2l}$'s} so that these priors are essentially non-informative.

\tr{
\begin{theorem} I). Assume that $\boldsymbol{b_m}, \tau_m^2$, and $\lambda_R^2$ are conditionally independent of $\zeta_{ijl}$ and $\lambda_l$, and $\sigma_{\epsilon}^2$ is conditionally independent of all other unknown parameters. Then the conditional posterior distributions of $\boldsymbol{b_m}, \tau_m^2,$ $\lambda_R^2$, and $\boldsymbol{c_r}$ are as follows:
\begin{align*}
f(\boldsymbol{b_m}|others) \propto \mbox{MVN}_v(\boldsymbol{\mu_{b_m}}, \boldsymbol{\Sigma_{b_m}}),
\end{align*}
where
\begin{center}
$\boldsymbol{\mu_{b_m}} = (\boldsymbol{\tilde{I}} + IJ\tau_m^2\boldsymbol{\Phi}(t)^T\boldsymbol{\Phi}(t))^{-1}[\tau_m^2\sum_{i}\sum_{j}(Y_{ij}(t) -\hat{Y}_{ij(-\boldsymbol{b_m})}(t))^T\boldsymbol{\Phi}(t)]^T,$ \\
\vspace{1em}
$\boldsymbol{\Sigma_{b_m}} = \sigma_{\epsilon}^2\tau_m^2(\boldsymbol{\tilde{I}} + IJ\tau_m^2\boldsymbol{\Phi}(t)^T\boldsymbol{\Phi}(t))^{-1},$
\end{center}
and $\boldsymbol{\tilde{I}}$ is a $v$ by $v$ identity matrix.
\begin{align*}
f(1 / \tau_m^2|others) \propto \mbox{IG}(v\lambda_R^2, \sqrt{\frac{v\lambda_R^2\sigma_{\epsilon}^2}{\boldsymbol{b_m}^T\boldsymbol{b_m}}}),
\end{align*}
and
\begin{align*}
f(\lambda_R^2|others) \propto \mbox{Gamma}(\alpha_{1R} + \frac{vp + p}{2}, \alpha_{2R} + \frac{v\sum_{m}\tau_m^2}{2}).
\end{align*}
The conditional posterior distribution of $\boldsymbol{c_r}$ (without penalization) is
\[
f(\boldsymbol{c_r}|others) \propto \mbox{MVN}_v(\boldsymbol{\mu_{c_r}}, \boldsymbol{\Sigma_{c_r}}),
\]
where
\begin{center}
$\boldsymbol{\mu_{c_r}} = (\Sigma_{\boldsymbol{c_r}}^{-1} + J(X_{ir}^c)^2\boldsymbol{\Phi}(t)^T\boldsymbol{\Phi}(t))^{-1}[\sum_{i}\sum_{j}(Y_{ij}(t) -\hat{Y}_{ij(-\boldsymbol{c_r})})^TX_{ir}^c\boldsymbol{\Phi}(t)]^T,$ \\
\vspace{1em}
$\boldsymbol{\Sigma_{c_r}} = \sigma_{\epsilon}^2(\Sigma_{\boldsymbol{c_r}}^{-1} + J(X_{ir}^c)^2\boldsymbol{\Phi}(t)^T\boldsymbol{\Phi}(t))^{-1}.$ 
\end{center}
II). Let $\boldsymbol{\zeta_{ij}} = (\zeta_{ij1}, ..., \zeta_{ijL})^T$ be an $L$-dimensional column vector of principal component scores, $\boldsymbol{\Lambda} = \mbox{diag}(\lambda_1, ..., \lambda_L)$ be a diagonal matrix containing eigenvalues, and $\boldsymbol{\Psi} = (\boldsymbol{\phi_1}(t),$ $\ldots, \boldsymbol{\phi_L}(t))_{K \times L}$ be a matrix containing eigenfunctions. The posterior distributions of $\boldsymbol{\zeta_{ij}}$, $\lambda_l$, and $\sigma_{\epsilon}^2$ are:
\begin{align*}
f(\boldsymbol{\zeta_{ij}}|others) \propto \mbox{MVN}_L(\boldsymbol{\mu_{\zeta_{ij}}}, \boldsymbol{\Sigma_{\zeta_{ij}}}),
\end{align*}
where
\begin{center}
$\boldsymbol{\mu_{\zeta_{ij}}} = (\boldsymbol{\Psi}^T\boldsymbol{\Psi} + \sigma_{\epsilon}^2(\boldsymbol{\Lambda})^{-1})^{-1}[(Y_{ij}(t) -\hat{Y}_{ij(-\boldsymbol{\zeta_{ij}})}(t))^T\boldsymbol{\Psi}]^T,$ \\
\vspace{1em}
$\boldsymbol{\Sigma_{\zeta_{ij}}} = \sigma_{\epsilon}^2(\boldsymbol{\Psi}^T\boldsymbol{\Psi} + \sigma_{\epsilon}^2(\boldsymbol{\Lambda})^{-1})^{-1},$ 
\end{center}
\begin{align*}
f(\lambda_l|others) \propto \mbox{IG}(\frac{1}{2}IJ + \alpha_{1l}, \frac{1}{2}\sum_i\sum_j \zeta_{ijl}^2 + \alpha_{2l}).
\end{align*}
And, for $\sigma_{\epsilon}^2$, we have:
\begin{align*}
f(\sigma_{\epsilon}^2|others) \propto \mbox{Scale-inv-}\chi^2 (IJK, \frac{\sum_{i}\sum_{j}(Y_{ij}(t) -\hat{Y}_{ij}(t))^T(Y_{ij}(t) -\hat{Y}_{ij}(t))}{IJK}),
\end{align*}
where $\mbox{Scale-inv-}\chi^2(\nu_1, \nu_2)$ is a scaled inverse chi-squared distribution with degrees of freedom $\nu_1$ and scale parameter $\nu_2$. See the supporting file for detailed derivations of the   theorem.
\end{theorem}
}

After the conditional posterior distributions of all of the unknown parameters are derived, we use Gibbs sampling to simulate these unknown parameters from their corresponding posterior distributions.  \tO{Plugging these estimates back to (\ref{bk}) and (\ref{Udecomp}), we will be able to estimate all curve/function related information: $\{\hat{\beta}_m(t)$, $\hat{C}_r(t)$; m = 1,...,p; r = 1,...,q\}, $\{\hat{Y}_{ij}(t); i=1,\ldots,I,j=1,\ldots,J\}$, $\{\hat{\phi}_l(t);l=1,\ldots,L\}$, and $\hat{\mu}(t)$. In addition, the proposed} \tr{HDFMM} approach also outputs the variable selection result, i.e., ranking the predictors according to their \tO{association strengths with the response, which are reflected through the $L_2$ norms of expansion coefficients (i.e., $||\boldsymbol{b_m}||;~m=1,\ldots,p$)}. Theoretically speaking, $||\boldsymbol{b_m}||=0$ if the $m^{th}$ predictor is not associated with the response. Therefore, the variable selection is performed by setting a threshold and removing those predictors whose rankings are below the threshold. The Bayesian information criterion (BIC) \citep{li2013selecting}, or other similar approaches, can be used to avoid the necessity of specifying a threshold for the magnitude of $\boldsymbol{b_m}$, which selects an optimal number of predictors.

\section{NUMERICAL STUDIES}
\subsection{Simulation studies}
In this section, we assess the performance of the \tr{HDFMM} method through multiple simulation studies. Following the standard procedure in the genome-wide association studies (GWAS) literature \citep{li2011bayesian, li2015}, the genotype of the $m^{th}$ marker for the $i^{th}$ subject is coded as additive and dominant dummy variables as follows

\begin{center}
$X_{im}^a = \begin{cases}
1, &  \mbox{if the genotype of the $m^{th}$ marker is AA,} \\ 0, & \mbox{if the genotype of the $m^{th}$ marker is Aa,} \\ -1, & \mbox{if the genotype of the $m^{th}$ marker is aa,}
\end{cases}$
\end{center}
\vspace{0.5em}

\begin{center}
$X_{im}^d = \begin{cases}
1, & \mbox{if the genotype of the $m^{th}$ marker is Aa,} \\ 0, & \mbox{if the genotype of the $m^{th}$ marker is AA or aa,}
\end{cases}$
\end{center}
where $i = 1, ..., I, m = 1, ..., p_1,$ and $p_1$ is the total number of markers. \tr{To adjust the need of GWAS application driven by the real data analysis, we slightly modify the model (\ref{main}) as the follows}
\begin{equation}
\begin{cases}
Y_{ij}(t) =  \boldsymbol{A}^T(t) \boldsymbol{X_i^a} + \boldsymbol{D}^T(t) \boldsymbol{X_i^d} +\boldsymbol{C}^T(t) \boldsymbol{X_i^c} + \sum_{l = 1}^{L} \zeta_{ijl}\phi_l(t)  + \epsilon_{ij}(t), \\
  \zeta_{ijl} \sim N(0, \lambda_l); \quad \epsilon_{ij}(t) \sim N(0, \sigma_{\epsilon}^2); \\
  i = 1, \ldots, I; ~j = 1, \ldots, J; l=1,\ldots,L; ~t \in [0,1].
\end{cases}
\label{full}
\end{equation}
\tO{where $\boldsymbol{X_i^a} = (X_{i1}^a, ..., X_{ip_1}^a)^T$ and $\boldsymbol{X_i^d} = (X_{i1}^d, ..., X_{ip_1}^d)^T$ are $p_1$-dimensional additive and dominant vectors containing all of the markers for subject $i$, and $\boldsymbol{A}(t) = [A_1(t),...,A_{p_1}(t)]$ and $\boldsymbol{D}(t) = [D_1(t),...,$ $D_{p_1}(t)]$ are their corresponding functional coefficients, respectively}. \ta{For the $m^{th}$ marker, $A_m(t)$ is interpreted as the average genotypic value when adding one major allele from the baseline, and $D_m(t)$ stands for the difference in the average genotypic value between heterozygote and homozygotes.}


We use the B-spline approach described in  (\ref{bk}) to approximate the two infinite-dimensional functional coefficient curves, $\boldsymbol{A}(t)$ and $\boldsymbol{D}(t)$, contained in (\ref{full}). Specifically, the functional additive and dominant coefficients of the $m^{th}$ marker is approximated as $A_m(t) = \boldsymbol{\Phi}(t) \boldsymbol{a_m}$ and $D_m(t) = \boldsymbol{\Phi}(t) \boldsymbol{d_m}, m = 1,...,p_1.$ Here $\boldsymbol{a_m} = (a_{m1}, ..., a_{mv})^T$ and $\boldsymbol{d_m} = (d_{m1}, ..., d_{mv})^T$ are their corresponding $v$-dimensional column vectors of the expansion coefficients. \ta{We apply the group Lasso to perform regularizations on $\boldsymbol{X_i^a}$ and $\boldsymbol{X_i^d}$ in model (\ref{full}) to address the ultrahigh dimensionality of the GWAS applications.} \tr{To reduce model complexity and computational cost, we use the same B-spline basis functions $\boldsymbol{\Phi}(t)$ for all the functions.}  

\subsubsection{Simulation designs}
In genome-wide association studies, the signal-to-noise ratio is usually very low. As a standard routine in simulations, we generate one covariate and 3,000 candidate markers ($p_1 = 3,000$ and $p=6,000$), of which only five markers are set to have influential additive and/or dominant functional genetic effects, and the majority of other markers are noise. We vary two levels of the number of subjects ($I = 100$ or $300$), and set the number of visits per subject to be $J = 5$ and the number of grids per curve to be $K = 50$. The genotypes of all of the markers (both the truly influential markers and the noise markers) are generated independently from a binomial distribution with a random minor allele frequency (MAF). That is, $X_m \sim \text{Binomial}(2, p_m)$, where $p_m$, the MAF of the $m^{th}$ marker, is simulated from $p_m \sim \mbox{Uniform}(0.1, 0.5)$. The continuous covariate is assumed to be standardized and is generated from a standard normal distribution $\mbox{N}(0, 1)$ as a result.

We perform two different simulation designs to test the robustness of the proposed \tr{HDFMM} model under various scenarios: 1) Simulation I: Directly assign the expansion coefficient vectors ($\boldsymbol{c_1}, \boldsymbol{a_1}, \boldsymbol{a_2}, \boldsymbol{a_3}, \boldsymbol{d_3}, \boldsymbol{a_4}, \boldsymbol{a_5}$) and B-spline functions for each influential marker and covariate, and then connect the response with the truly influential predictors directly using (\ref{full}). We set five basis functions ($v = 5$), in which cubic splines and default knots of 0, 0.5 and 1 are used. 2) Simulation II: Indirectly assign specific functions to the functional coefficient ($C_1(t), A_1(t), A_2(t), A_3(t), D_3(t), A_4(t), A_5(t)$) of each influential marker, and then connect the response with truly influential predictors using (\ref{full}). For both simulations, we set the $3^{rd}$ marker to have both additive and dominant effects and the other four markers to have only additive effects. See Table \ref{tab3.1} for the detailed values assigned to the expansion coefficient vectors (Simulation I) and to the functional coefficients (simulation II) for the five influential markers, and to the covariate used in the simulation studies.

\begin{table}
\begin{center}
\caption{\label{tab3.1} Coefficients and functions used in Simulation I and Simulation II}
\begin{tabular}{ c | c | c c c c c}
\hline
Simulation & Parameter & \multicolumn{5}{c}{Expansion Coefficients} \\ \hline
&$\boldsymbol{c_1}$ & 1 & 0 & -4 & -1 & 4 \\
&$\boldsymbol{a_1}$ & 4 & 1 & 3 & 4 & 2 \\
&$\boldsymbol{a_2}$ & 3 & 3 & 0 & -4 & 3 \\
I &$\boldsymbol{a_3}$ & 2 & 5 & 0 & 0 & 0 \\
&$\boldsymbol{d_3}$ & 1 & 1 & 5 & 1 & 1 \\
&$\boldsymbol{a_4}$ & 4 & 3 & 0 & 1 & 3 \\
&$\boldsymbol{a_5}$ & 1 & 1 & 1 & 1 & -5 \\ \hline \hline
Simulation & Parameter & \multicolumn{5}{c}{Functional Coefficients} \\ \hline
& $C_1(t)$ & \multicolumn{5}{c}{$t / 10$} \\
&$A_1(t)$ & \multicolumn{5}{c}{$10\sqrt{t}$} \\
&$A_2(t)$ & \multicolumn{5}{c}{$\mbox{exp}(2t)$} \\
II &$A_3(t)$ & \multicolumn{5}{c}{$5t^2$} \\
&$D_3(t)$ & \multicolumn{5}{c}{$t^3 / 3$} \\
&$A_4(t)$ & \multicolumn{5}{c}{$1 - 2^t$} \\
&$A_5(t)$ & \multicolumn{5}{c}{$10t$} \\
\hline
\end{tabular}
\end{center}
\end{table}

For the nested subject/visit-specific variance component $U_{ij}(t)$ \ta{(set to be the same for both simulations)}, we set $L=4$, with eigenvalues $\lambda_l's$ set to be 1, 0.9, 0.6, and 0.5. Then the subject/visit-specific principal component scores $\{\zeta_{ijl}\}$ are generated from their corresponding eigenvalues, $\zeta_{ijl} \sim \mbox{N}(0, \lambda_l)$. The corresponding \tO{eigenfunctions} are set as: 
\begin{center}
$\{\phi_1(t), \phi_2(t), \phi_3(t), \phi_4(t)\} = \{\sqrt{2} \mbox{sin}(4\pi t), \sqrt{2} \mbox{cos}(4\pi t), \sqrt{2} \mbox{sin}(8\pi t), \sqrt{2} \mbox{cos}(8\pi t)\}$. 
\end{center}
Finally, the variance of the measurement error $\epsilon_{ij}(t)$ is set as $\sigma_{\epsilon}^2=0$ (no noise) or $\sigma_{\epsilon}^2=1$ (moderately noisy), which have the same magnitudes of errors as the simulation studies of \citet{crainiceanu2009}. 

\subsubsection{Simulation results}
Three criteria are used to evaluate the performance of the \tr{HDFMM} method \citep{li2012feature, carlsen2016exploiting}:
\begin{itemize}
\item \ta{`Individual power' $p_m, m=1,\ldots,5$: the percentage of each influential marker being successfully selected across all simulation replicates.}
\item `Strict power': the \tO{proportion of simultaneously identifying all five influential markers} across all simulation replicates;
\item `Type I error': the average proportion of mistakenly identifying non-influential markers across all simulation replicates.
\end{itemize}

\begin{table}[p]
\centering
\begin{center}
\caption{\label{tab2} The overall performance of the \tr{HDFMM} model through simulation studies.}
\begin{tabular}{c | c c c c}
\hline
Simulation & $\sigma_{\epsilon}$ & $I$ & Strict Power & Type I Error \\ \hline
\multirow{4}{*}{I} & 0 & 100 & 96\% & 0.04\% \\
& 0 & 300 & 100\% & 0.03\% \\
& 1 & 100 & 86\% & 0.56\% \\
& 1 & 300 & 100\% & 0.28\% \\
\hline \hline
\multirow{4}{*}{II} & 0 & 100 & 78\% & 0.05\% \\
& 0 & 300 & 100\% & 0.01\% \\
& 1 & 100 & 44\% & 0.01\% \\
& 1 & 300 & 82\% & 0.01\% \\
\hline
\end{tabular}
\end{center}
\end{table}
\begin{table}
\centering
\begin{center}
\caption{\label{tab3} The individual powers of the \tr{HDFMM} model through simulation studies.}
\begin{tabular}{c | c c c c c c c c}
\hline
Simulation & $\sigma_{\epsilon}$ & $I$ & & $p_1$ & $p_2$ & $p_3$ & $p_4$ & $p_5$ \\ \hline
\multirow{4}{*}{I} & 0 & 100 & & 100\% & 100\% & 100\% & 100\% & 80\% \\
& 0 & 300 & & 100\% & 100\% & 100\% & 100\% & 100\% \\
& 1 & 100 & & 100\% & 100\% & 100\% & 80\% & 50\% \\
& 1 & 300 & & 100\% & 100\% & 100\% & 100\% & 100\% \\
\hline \hline
\multirow{4}{*}{II} & 0 & 100 & & 100\% & 100\% & 80\% & 80\% & 30\% \\
& 0 & 300 & & 100\% & 100\% & 100\% & 100\% & 100\% \\
& 1 & 100 & & 90\% & 90\% & 30\% & 10\% & 1\% \\
& 1 & 300 & & 100\% & 100\% & 100\% & 80\% & 30\% \\
\hline
\end{tabular}
\end{center}
\end{table}

As shown in Tables~\ref{tab2} and \ref{tab3}, the simulation results of the \tr{HDFMM} method are very promising. For Simulation I, 100\% `powers' can be achieved and `type I error rates' are well-controlled when the sample size is at least 300. In fact, it is notable that the \tr{HDFMM} model can successfully detect all of the five influential markers at all 100 replicates given the need to handle 1500 curves (with 50 time points for each curve) and 6000 markers whose majority components are noise. When we increase the variation of the noise, the power decreases to 86\% if the sample size is too small ($n=100$), but it regains 100\% when the sample size is back to $n=300$. It is to be expected that the `type I error rate' given higher variation is larger than when lower variation is applied (0.28\% versus 0.03\%), but they are still lower than 5\%.

\begin{figure}[t]
\centering
\makebox{\includegraphics[width=250pt,height=350pt]{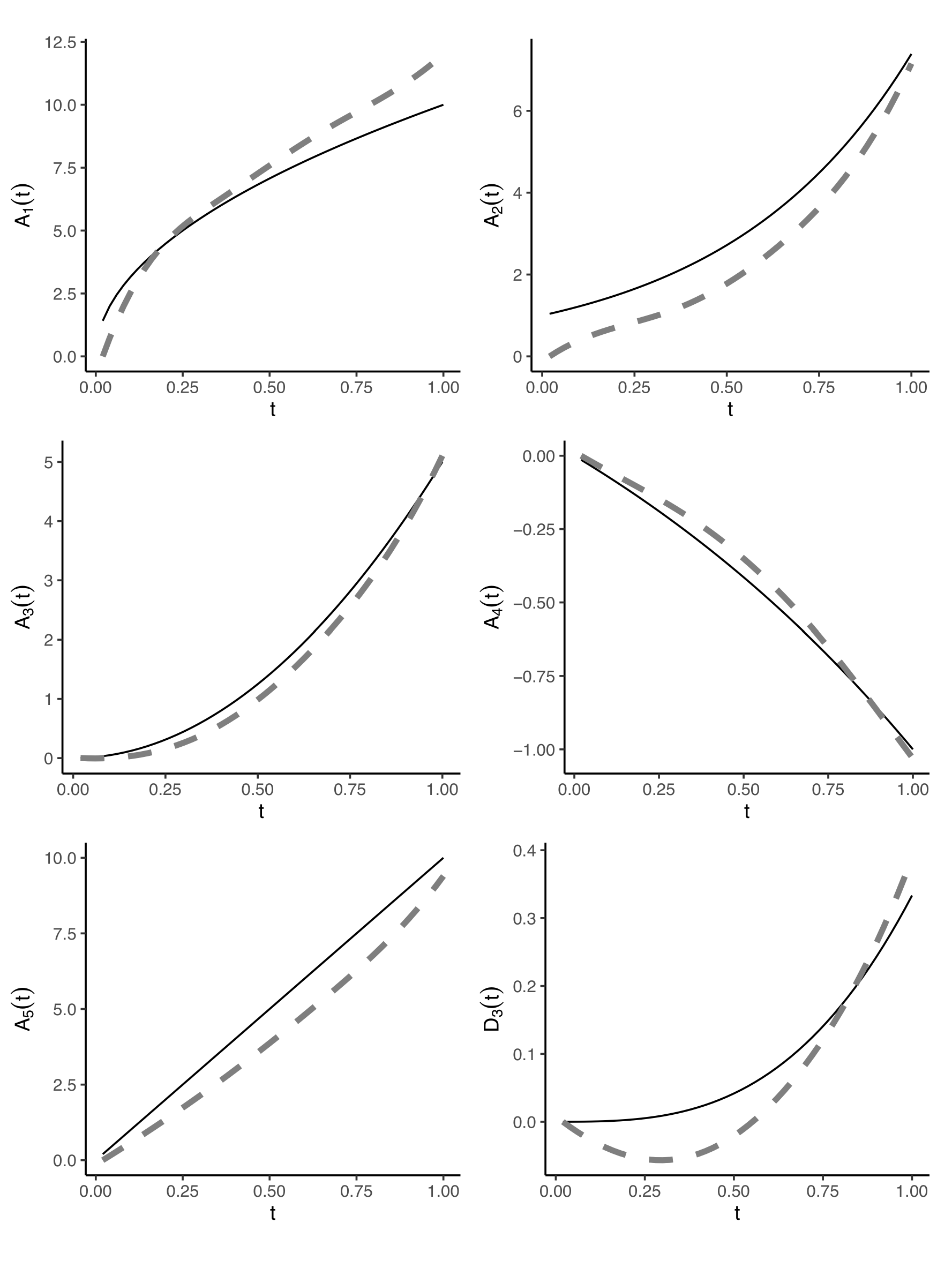}} 
\caption{\label{fig3.0a}
The true and estimated functional effects of influential markers in a randomly selected replicate of Simulation II generated under the setting of $I = 300; \sigma_{\epsilon} = 1$. The true functional effects are plotted in solid lines, and the estimated functional effects are plotted in dashed lines.}
\end{figure}

\begin{figure}[h]
\centering
\hspace{-5em}
\makebox{\includegraphics[width=250pt,height=230pt]{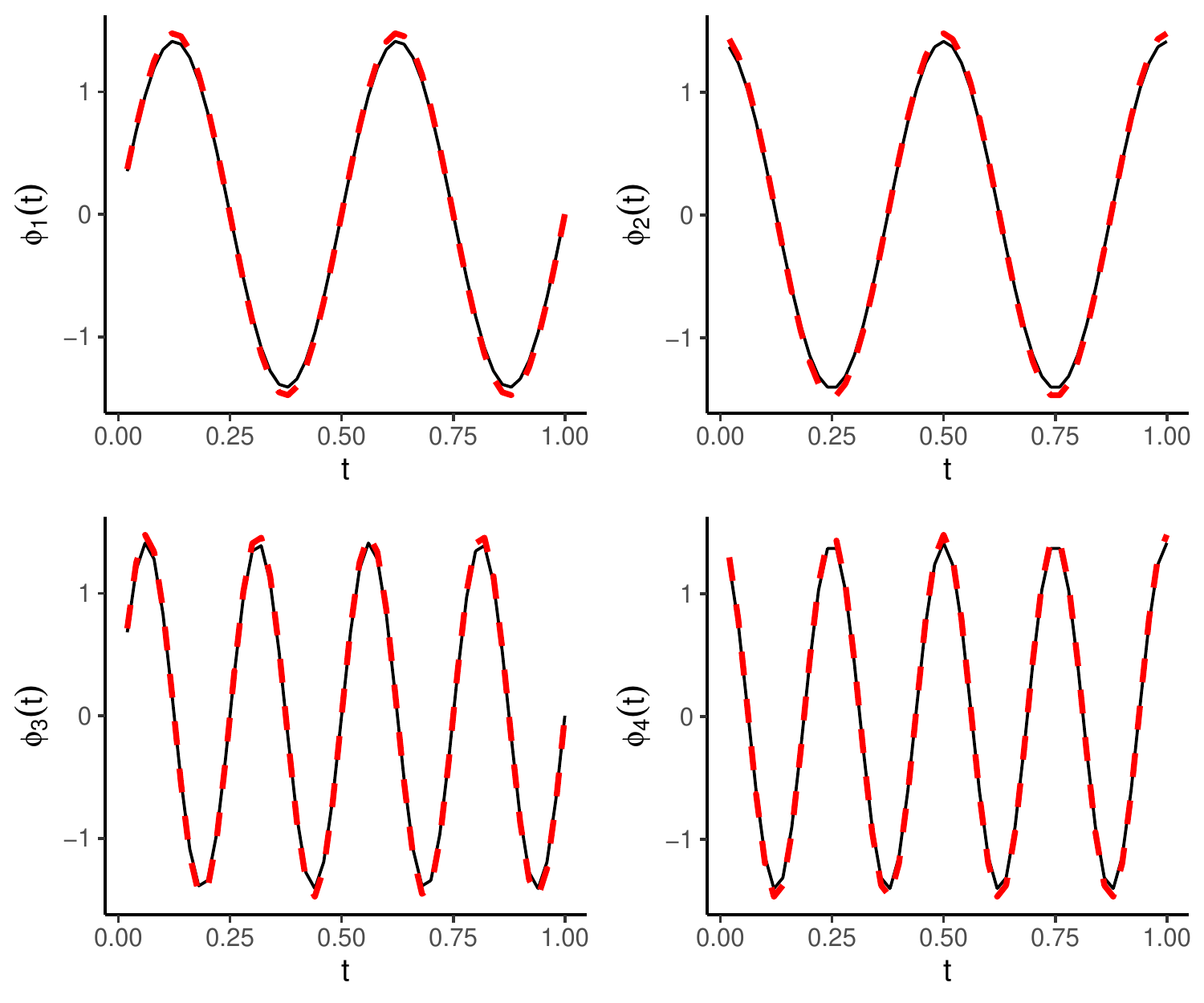}} 
\vspace{-1em}
\caption{\label{fig3.0b}
The true and estimated eigenfunctions of the subject/visit-specific residual ($U_{ij}(t)$) in a randomly selected replicate of Simulation II generated under the setting of $I = 300; \sigma_{\epsilon} = 1$. The true eigenfunctions are plotted in solid lines, and the estimated eigenfunctions are plotted in dashed lines.}
\end{figure}

In Simulation II, we increased the level of difficulty by connecting the response and predictors indirectly. So, we are not surprised to see that the results of Simulation II are lower than those of Simulation I. But, `powers' of 100\% and 82\% and `type I error rates' of 0.01\% are still promising if the sample size is at least 300 under this harder scenario. In addition to the `power' results, we also compare the estimated functional coefficients ($\hat{A}_1(t), \hat{A}_2(t), \hat{A}_3(t), \hat{D}_3(t), \hat{A}_4(t), \hat{A}_5(t)$) to those of true functional coefficients ($A_1(t), A_2(t), A_3(t)$, $D_3(t), A_4(t), A_5(t)$) for each of the truly influential markers. The estimated functional coefficients in Simulation II are obtained by averaging results across all of the replicates in the setting of $I = 300$ and $\sigma_{\epsilon} = 1$. The true functional coefficients are plotted in solid lines and the estimated functional coefficients are plotted in dashed lines (see Figure \ref{fig3.0a}). As we observed, the estimated functional coefficient for each of the influential markers is notably above the zero line. In addition, most estimated functional trends precisely capture the true functions specified in Table \ref{tab3.1}.

To evaluate the effectiveness of the \tr{HDFMM} approach on estimating the subject/visit-specific variance component $U_{ij}$, we also compare the estimated eigenfunctions of $U_{ij}(t)$ (i.e., $\hat{\phi}_1(t), \hat{\phi}_2(t), \hat{\phi}_3(t), \hat{\phi}_4(t)$) to the true eigenfunctions (i.e., $\phi_1(t), \phi_2(t), \phi_3(t), \phi_4(t)$). As shown in Figure \ref{fig3.0b}, the estimated eigenfunctions (plotted in dashed lines) effectively and accurately approximate the true eigenfunctions (plotted in solid lines).

\subsection{Real data analysis}
\begin{figure}[t]
\centering
\makebox{\includegraphics[width=380pt, height=150pt]{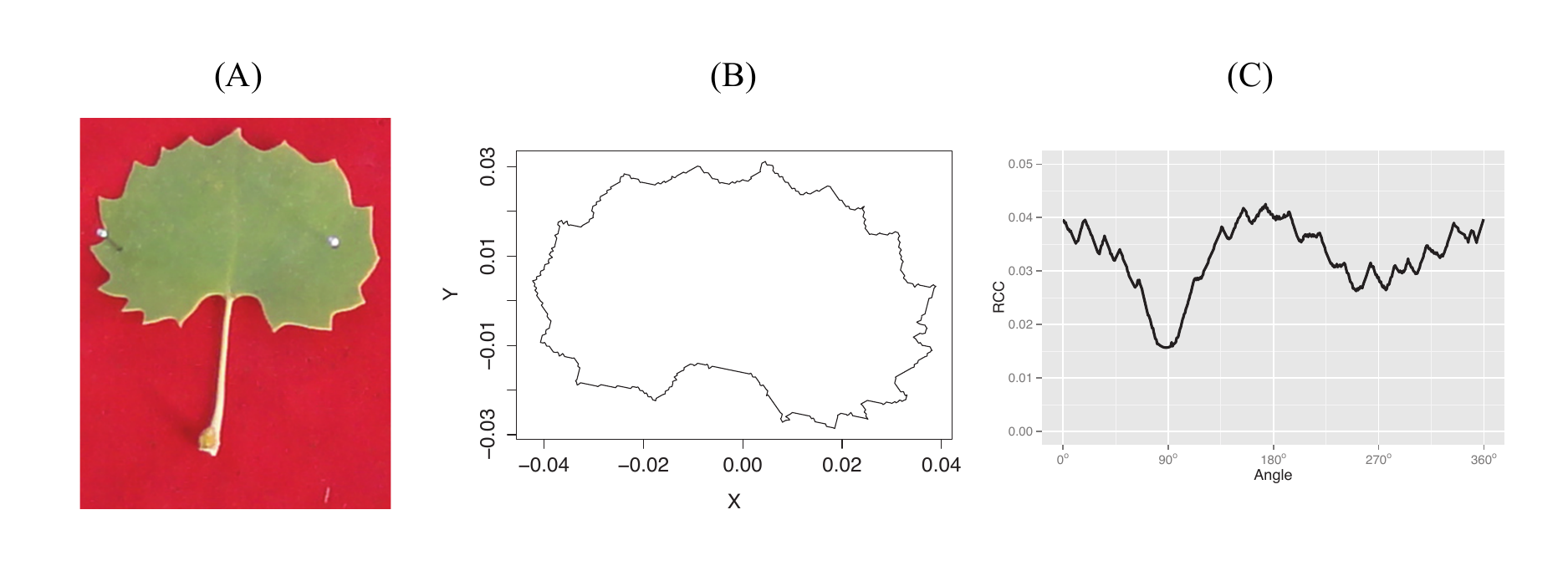}} 
\vspace{-1em}
\caption{\label{raw}
The process of transforming leaf shapes into functional curves. (A) The original picture of a \emph{Populus euphratica} leaf; (B) The leaf shape boundary is mapped into x-y coordinates; (C) The x-y coordinates is transformed into a functional radii curve.}
\end{figure}

We now analyze the leaf shapes of a natural population of 421 \emph{Populus euphratica} plants. \tr{The leaf shapes of \emph{Populus euphratica} are polymorphic with complex boundaries, so we employ a high dimensional directional radii curve with $910 \times 1$ dimension to accurately represent each leaf shape (see Figure\ref{raw}(A) for an example of \emph{Populus euphratica} leaf). After fixing a starting point and an end point, the x-y coordinates of all points on the boundary of each shape are recorded one by one (see Figure\ref{raw}(B)). Then the radii connecting each point on the boundary to the centroid are computed and formed a radii curve (see Figure\ref{raw}(C)). Then we perform the alignment to filter out the variations caused by pose (translation, scale, and rotation), as detailed in \citet{kong2007}. The radii curve is a standard functional curve measured on dense and regular grids.}

\begin{figure}
\centering
        \includegraphics[width=360pt, height=250pt]{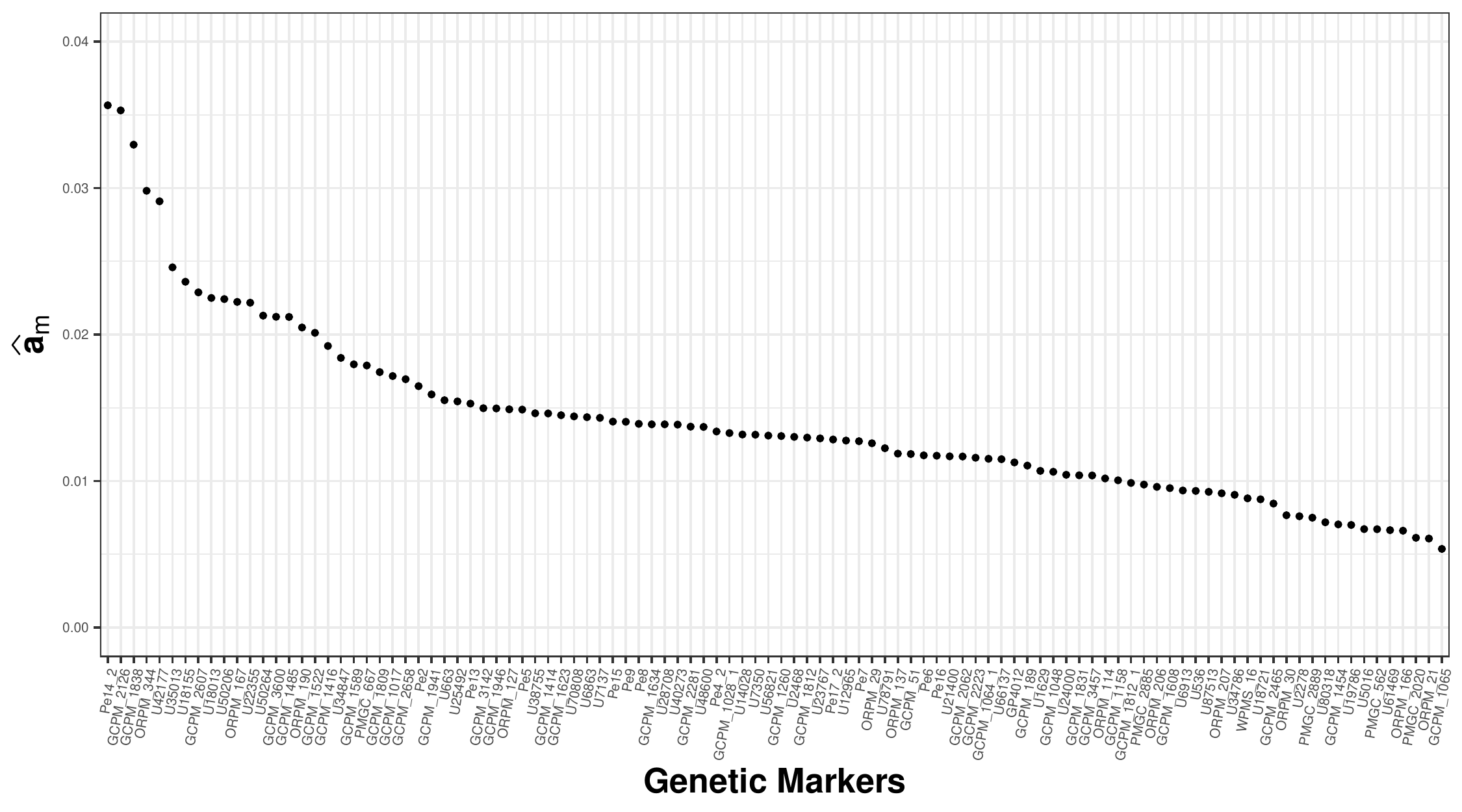}
        \vspace{2em}
        
        \includegraphics[width=360pt, height=250pt]{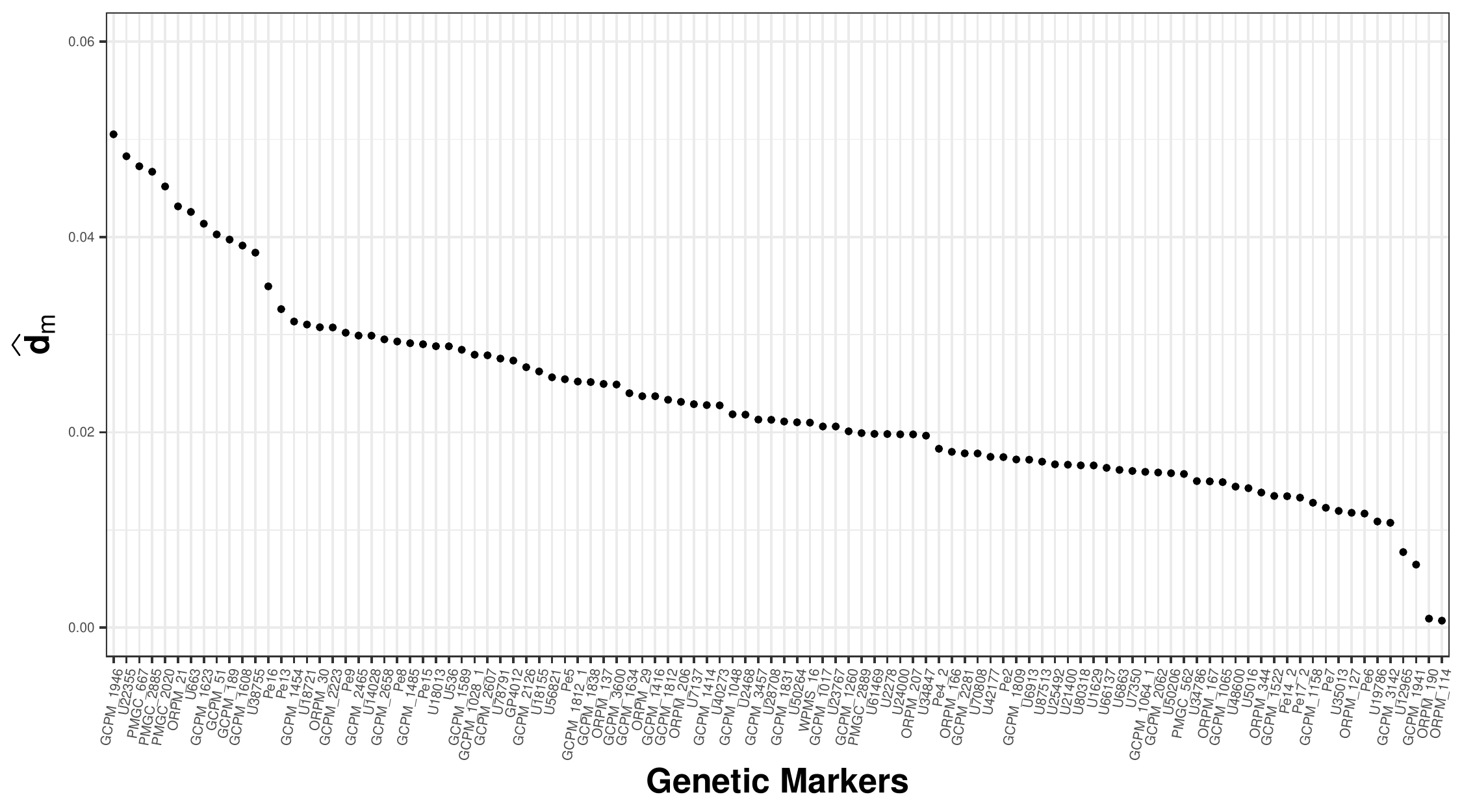}
    \caption{\label{l2rank} \trr{The $l_2$ norm of the estimated additive coefficients (top panel) and estimated dominant coefficients (bottom panel) for each of the 104 markers, respectively, ranking from the highest to the lowest.}}
\end{figure}

\begin{figure}
\centering
        \includegraphics[width=360pt, height=200pt]{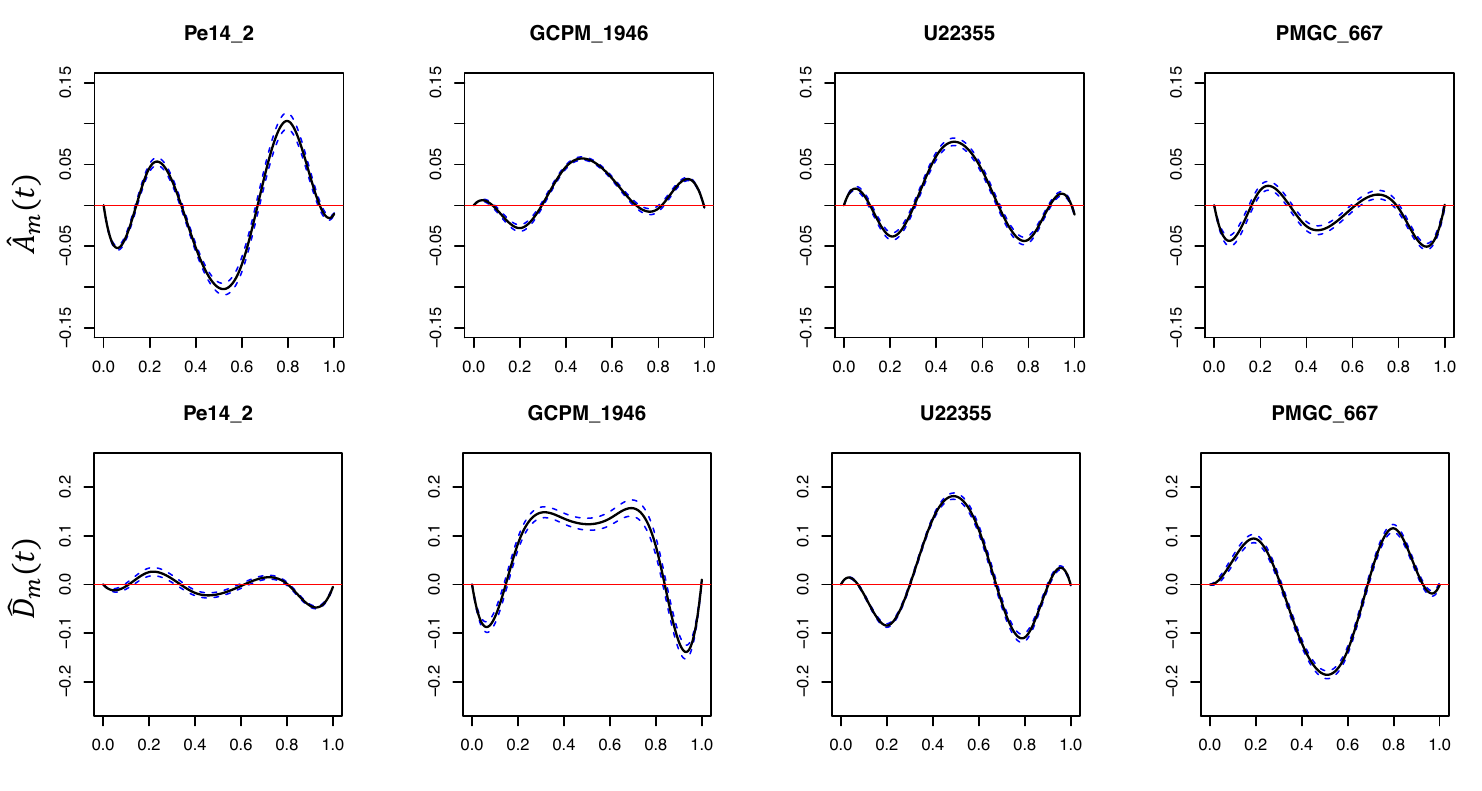}
    \caption{\label{CI} \trr{The estimated time-varying coefficients for the additive effect $\hat{A}_m(t)$ (top panel) and dominant effect $\hat{D}_m(t)$ (bottom panel) in solid black line, along with their corresponding 95\% confidence bands in dashed blue line. The red zero line indicates no functional effects.}}
\end{figure}

\begin{figure}[t]
\centering
\makebox{\includegraphics[width=400pt, height=300pt]{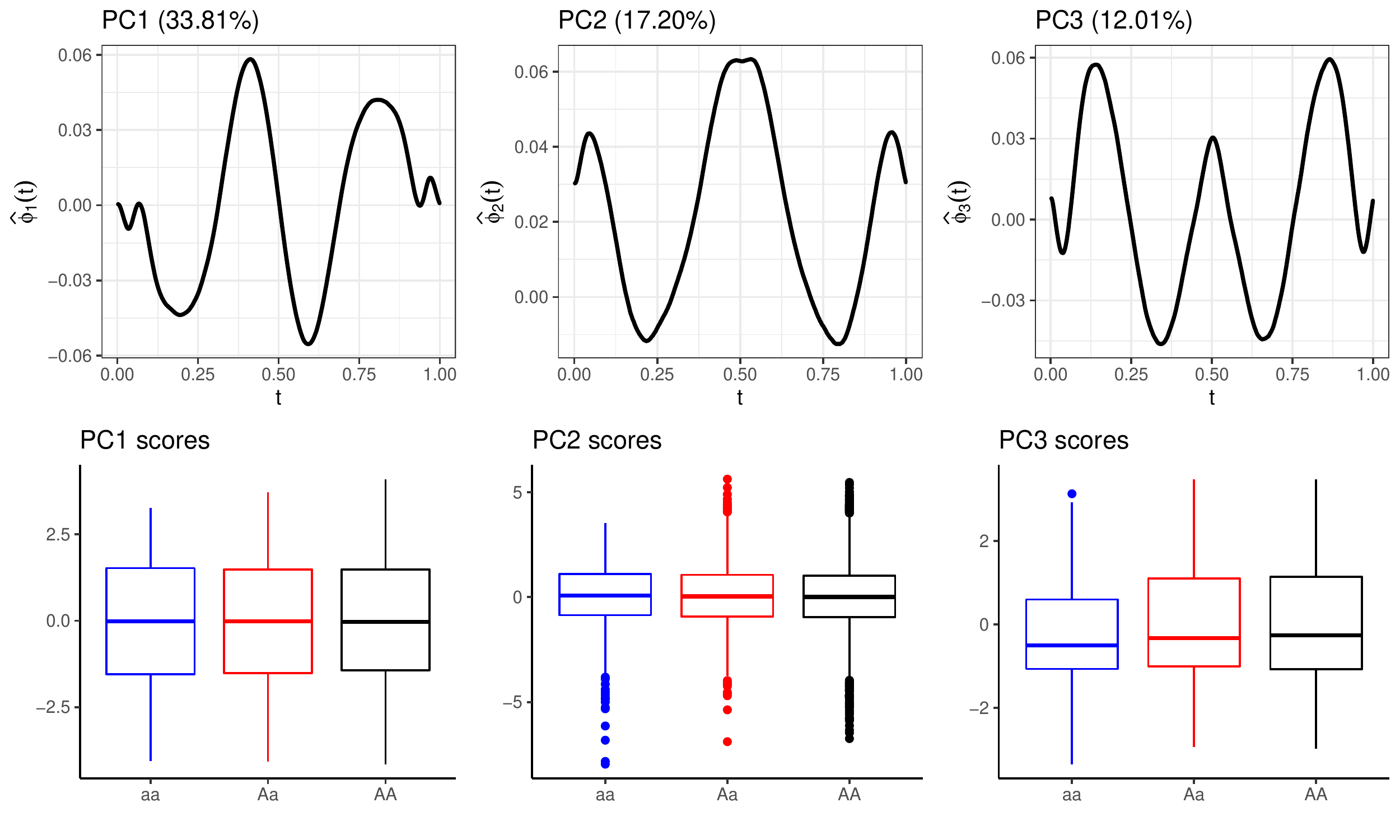}} 
\vspace{-1em}
\caption{\label{pcscore}
\trr{The first three estimated eigenfunctions $\{\hat{\phi}_l(t),~l=1,2,3\}$ (top panel) and corresponding PC scores $\{\hat{\zeta}_{ijl},~l=1,2,3\}$ for $\hat{U}_{ij}(t)$ (bottom panel) estimated from the model (\ref{full}). The PC scores are divided into three boxplots with three different colors according to three genotypes of marker \emph{Pe14\_2} (\emph{AA} in black, \emph{Aa} in red, and \emph{aa} in blue).}}
\end{figure}

Twenty-five leaves were randomly collected from each tree to balance off the environmental factors. However, computing an average of twenty-five leaves from each plant and then fitting a single-level model would result in a significant loss of descriptive accuracy in shape because it will smooth out several sharp and irregular teeth details.  To improve the accuracy, we apply the \tr{HDFMM} method to directly model the twenty-five repeatedly measured and hence correlated curves for each of the 421 trees. In addition to the 10,525 shape curves in total, the data also include 104 genotyped genetic markers (i.e., $I = 421, J = 25, T=910, p_1 = 104$), which represent a challenging modeling task.

We decide to keep the top fifteen principal components for the visit/subject-specific variance term, $U_{ij}(t)$, based on the rules described in the methodology section. The functional coefficients are approximated by seven basis functions ($v = 7$), in which cubic splines and five equidistant knots are used.  For each MCMC chain, we perform 1,000 burn-in iterations followed by 1,000 updating iterations, and the unknown parameters are estimated by taking the average of each of the 1,000 updating iterations. The confidence band (CB) for each additive and dominant functional coefficient is estimated from the 1,000 updating iterations of corresponding expansion coefficients, inspired by \cite{jones2006}, but transformed to curves by multiplying the basis functions. 

The $l_2$ norms of the estimated additive and dominant coefficients of the 104 markers (i.e., $||\boldsymbol{a_m}||$ and $||\boldsymbol{d_m}||$) are shown in Figure \ref{l2rank}, ranking from the largest to the smallest, respectively. We determine that a genetic marker is important if either its additive effect or dominant effect or both is ranked in the top list. The BIC criterion is employed to determine the optimal number of markers to keep, i.e., the cutoff value, based on the model structure (\ref{full}). Specifically, we start with the top ranked marker and then gradually involve only one more marker at a time in an order suggested by Figure \ref{l2rank} until reach the full model containing all of the markers. Then the BIC value is computed and compared for each candidate model under consideration. At the end we detect four markers, $\{ \emph{Pe14\_2, GCPM\_1946, U22355, PMGC\_667} \}$, whose BIC value achieves the minimum among all candidate models.  

 The estimated time-varying or location-varying functional coefficient trajectories of the additive effect $\hat{A}_m(t)$ and dominant effect $\hat{D}_m(t)$ for each of these four selected important markers are demonstrated in Figure \ref{CI}, respectively, along with their corresponding 95\% confidence band.  Almost all functional coefficients and their corresponding CBs of these four selected markers show strong fluctuating dynamic trends that significantly deviate from the horizontal axis, empirically indicating the necessity of modeling the coefficients of functional data as a curve rather than as a scalar. These confidence bands in Figure \ref{CI} do not cover zero in any interval; it confirms the biological importance of these four selected markers in regulating leaf shape variation. In addition, we observe that the confidence bands are narrow, which indicates the small variations among the 1,000 updating MCMC interactions. 
 
The top panel of Figure~\ref{CI} demonstrates that the estimated time-varying functional coefficients of the selected markers can be mostly negative (e.g., \emph{PMGC\_667}), be mostly positive (e.g., \emph{GCPM\_1946}), or have varying signs at different time or location intervals (e.g., \emph{Pe14$\_$2, U22355}). Recall that the coding of an additive dummy variable is 1 (AA), 0 (Aa), and -1 (aa). Hence an additive genetic effect can be interpreted as the average change in response value when adding or removing one major allele (\emph{A}) from the baseline genotype (\emph{Aa}). A positive value of the additive coefficient of a specific marker indicates that the individuals who carry genotype \emph{AA} have larger expected response values and who carry genotype \emph{aa} have smaller expected response values than those carrying genotype \emph{Aa}, assuming all of the other markers remain the same. Similarly, the situation for a negative value of the additive coefficient is opposite. Note that what a functional coefficient demonstrates is a curve instead of a single value; therefore, aforementioned signs of a functional coefficient should be interpreted on a time- or location-interval. Specifically, \emph{Pe14$\_$2} has positive additive coefficient on intervals [0.15,0.375] and [0.7,0.85] (corresponding to $[55^{o},135^{o}]$ and $[250^{o},305^{o}]$ ) and remain negative on the remaining intervals. Since the measurement for shape is the radius, larger values indicate that the shape outline outwards from the center, while small values retract towards the center.   

In Figure \ref{pcscore}, we also demonstrate the first three estimated eigenfunctions $\{\hat{\phi}_l(t),~l=1,2,3\}$ and corresponding PC scores $\{\hat{\zeta}_{ijl},~l=1,2,3\}$ for $\hat{U}_{ij}(t)$ estimated from model (\ref{full}). The PC scores are divided into three boxplots with three different colors according to three genotypes of marker \emph{Pe14\_2} (\emph{AA} in black, \emph{Aa} in red, and \emph{aa} in blue). We have noticed that the averages of these three boxplots are similar because no genetic information should be remaining in the random term if the genetic markers are already modeled as fixed effects in model (\ref{full}).

\section{CONCLUSION AND DISCUSSION}
\tf{The HDFMM model provides an alternative option to current functional mixed effect model literature with the focus on high dimensional bilevel functional data when the number of predictors is usually larger than the number of subjects ($p>n$). Its far-reaching application comes from its nonparametric nature in modeling random effect terms and flexibility in handling complex situations with high dimensions, correlations, and heterogeneities in various forms. }

The \tr{HDFMM} model inherits the effectiveness of the linear mixed effect model in accounting for relatedness among samples and in controlling for confounding factors \citep{zhang2010mixed}.  In addition to high-dimensional predictors, it also allows for covariates serving control purposes, such as age, population structure, treatment, etc.
 
We consider only equidistant knots in this article; however, modifying a B-spline system to non-equal distance knots, without changing other parts of the \tr{HDFMM} method, is subject to minor changes. In addition, other approaches used to smooth and approximate functions, such as local polynomial methods \citep{fan1995local}, smoothing spline methods \citep{eilers1996flexible}, and Legendre polynomials \citep{li2015}, are feasible alternatives that do not require changing the model structure of the \tr{HDFMM} method. In addition to the FACE algorithm that we utilize to smooth the covariance matrix, a bivariate P-spline smoother is also applicable, as suggested by \cite{yao2006} and \cite{di2009multilevel}. The scope of this article mainly focuses on equally-spaced regular measurements for functional data, and we will leave the adjustment of \tr{HDFMM} to random or irregularly-spaced measuring points for longitudinal data to future works. For example, incorporating the skills of time-warping adjustments \citep{yi1998efficient} or treating the time points as an extra random variable \citep{yao2005, mulleryao2} or adopting the PACE algorithm proposed by \cite{yao2005}.

\section{Appendix}

\section*{Proof of \emph{Theorem 2.1.}}
\noindent \textbf{\emph{Part I).}} All the conjugate priors have standard-form conditional posterior distributions. Assuming that $\boldsymbol{b_m}, \tau_m^2$, and $\lambda_R^2$ are conditionally independent of $\zeta_{ijl}$ and $\lambda_l$, and $\sigma_{\epsilon}^2$ is conditionally independent of all other unknown parameters, the joint posterior distribution containing all unknown parameters is
\begin{align*}
f&(\boldsymbol{c_r}, \boldsymbol{b_m}, \tau_m^2, \lambda_R^2, \zeta_{ijl}, \lambda_l, \sigma_{\epsilon}^2|\ta{\boldsymbol{y}}) \\
& \propto f(\ta{\boldsymbol{y}}|\ta{\text{all parameters}})f(\sigma_{\epsilon}^2)\prod_{r=1}^{q}f(\boldsymbol{c_r}) \\
& \quad \times\prod_{m=1}^{p}f(\boldsymbol{b_m}|\tau_m^2)f(\tau_m^2|\lambda_R^2)f(\lambda_R^2) \\ 
& \quad \times\prod_{i=1}^{I}\prod_{j=1}^{J}\prod_{l=1}^{L}f(\zeta_{ijl}|\lambda_l)\prod_{l=1}^{L}f(\lambda_l).
\end{align*}

The conditional posterior distributions of $\boldsymbol{b_m}, \tau_m^2,$ and $\lambda_R^2$ can be derived as follows:
\begin{align*}
f&(\boldsymbol{b_m}|others) \\
&\propto f(\ta{\boldsymbol{y}}|others)f(\boldsymbol{b_m}|\tau_m^2) \\
&\propto \mbox{exp}[-\frac{1}{2\sigma_{\epsilon}^2} (\frac{\boldsymbol{b_m}^T\boldsymbol{b_m}}{\tau_m^2} + \ta{\sum_{i}\sum_{j} (Y_{ij}(t) -\hat{Y}_{ij}(t))^T(Y_{ij}(t) -\hat{Y}_{ij}(t)))]} \\
&\propto \mbox{exp}[-\frac{1}{2\sigma_{\epsilon}^2\tau_m^2}(\boldsymbol{b_m}^T(\boldsymbol{\tilde{I}} + IJ\tau_m^2\ta{\boldsymbol{\Phi}(t)^T\boldsymbol{\Phi}(t))}\boldsymbol{b_m} \\
&- \ta{2\tau_m^2\sum_{i}\sum_{j}(Y_{ij}(t) -\hat{Y}_{ij(-\boldsymbol{b_m})}(t))^T\boldsymbol{\Phi}(t)\boldsymbol{b_m})]} \\
&\propto \mbox{MVN}_v(\boldsymbol{\mu_{b_m}}, \boldsymbol{\Sigma_{b_m}}),
\end{align*}
where
\begin{center}
$\boldsymbol{\mu_{b_m}} = (\boldsymbol{\tilde{I}} + IJ\tau_m^2\ta{\boldsymbol{\Phi}(t)^T\boldsymbol{\Phi}(t)})^{-1}\ta{[\tau_m^2\sum_{i}\sum_{j}(Y_{ij}(t) -\hat{Y}_{ij(-\boldsymbol{b_m})}(t))^T\boldsymbol{\Phi}(t)]^T,}$ \\
\vspace{1em}
$\boldsymbol{\Sigma_{b_m}} = \sigma_{\epsilon}^2\tau_m^2(\boldsymbol{\tilde{I}} + IJ\tau_m^2\ta{\boldsymbol{\Phi}(t)^T\boldsymbol{\Phi}(t)})^{-1},$
\end{center}
and $\boldsymbol{\tilde{I}}$ is a $v$ by $v$ identity matrix.
\begin{align*}
f&(\tau_m^2|others) \\
&\propto f(\boldsymbol{b_m}|\tau_m^2)f(\tau_m^2|\lambda_R^2) \\
\vspace{2em}
&\propto (\tau_m^2)^{-\frac{1}{2}}\mbox{exp}[-\frac{1}{2}(\frac{\boldsymbol{b_m}^T\boldsymbol{b_m}}{\tau_m^2\sigma_{\epsilon}^2} + v\lambda_R^2\tau_m^2)].
\end{align*}
\ta{It implies} $f(1 / \tau_m^2|others) \propto \mbox{IG}(v\lambda_R^2, \sqrt{\frac{v\lambda_R^2\sigma_{\epsilon}^2}{\boldsymbol{b_m}^T\boldsymbol{b_m}}})$.
\begin{align*}
f&(\lambda_R^2|others) \\
&\propto \prod_{m=1}{p}f(\tau_m^2|\lambda_R^2)f(\lambda_R^2) \\
&\propto (\lambda_R^2)^{\alpha_{1R} - 1} \mbox{exp}(-\alpha_{2R} \lambda_R^2) \times (\frac{v\lambda_R^2}{2})^{\frac{vp + p}{2}}\mbox{exp}[-\lambda_R^2\frac{v\sum_{m}\tau_m^2}{2}] \\
&\propto \mbox{Gamma}(\alpha_{1R} + \frac{vp + p}{2}, \alpha_{2R} + \frac{v\sum_{m}\tau_m^2}{2}).
\end{align*}

The conditional posterior distribution of $\boldsymbol{c_r}$ (without penalization) is
\[
f(\boldsymbol{c_r}|others) \propto \mbox{MVN}_v(\boldsymbol{\mu_{c_r}}, \boldsymbol{\Sigma_{c_r}}),
\]
where
\begin{center}
$\boldsymbol{\mu_{c_r}} = (\Sigma_{\boldsymbol{c_r}}^{-1} + J(X_{ir}^c)^2\boldsymbol{\Phi}(t)^T\boldsymbol{\Phi}(t))^{-1}[\sum_{i}\sum_{j}(\ta{Y_{ij}(t) -\hat{Y}_{ij(-\boldsymbol{c_r})})^TX_{ir}^c\boldsymbol{\Phi}(t)}]^T,$ \\
\vspace{1em}
$\boldsymbol{\Sigma_{c_r}} = \sigma_{\epsilon}^2(\Sigma_{\boldsymbol{c_r}}^{-1} + J(X_{ir}^c)^2\ta{\boldsymbol{\Phi}(t)^T\boldsymbol{\Phi}(t)})^{-1}.$ 
\end{center}

\vspace{2em}

\noindent \textbf{\emph{Part II).}} Let $\boldsymbol{\zeta_{ij}} = (\zeta_{ij1}, ..., \zeta_{ijL})^T$ be an $L$-dimensional column vector of principal component scores, $\boldsymbol{\Lambda} = \mbox{diag}(\lambda_1, ..., \lambda_L)$ be a diagonal matrix containing eigenvalues, and \ta{$\boldsymbol{\Psi} = (\boldsymbol{\phi_1}(t), \ldots, \boldsymbol{\phi_L}(t))_{K \times L}$} be a matrix containing eigenfunctions. The posterior distributions of $\boldsymbol{\zeta_{ij}}$ and $\lambda_l$ are:
\begin{align*}
f&(\boldsymbol{\zeta_{ij}}|others) \\
\vspace{2em}
&\propto f(\ta{\boldsymbol{y}}|others)f(\boldsymbol{\zeta_{ij}}|\boldsymbol{\Lambda}) \\
&\propto \mbox{exp}[-\frac{\ta{(Y_{ij}(t) -\hat{Y}_{ij}(t))^T(Y_{ij}(t) -\hat{Y}_{ij}(t))}}{2\sigma_{\epsilon}^2} - \frac{1}{2} \boldsymbol{\zeta_{ij}}^T(\boldsymbol{\Lambda})^{-1}\boldsymbol{\zeta_{ij}}] \\
&\propto \mbox{exp}[-\frac{1}{2\sigma_{\epsilon}^2} (\boldsymbol{\zeta_{ij}}^T(\boldsymbol{\Psi}^T\boldsymbol{\Psi} + \sigma_{\epsilon}^2(\boldsymbol{\Lambda})^{-1})\boldsymbol{\zeta_{ij}} - 2(\ta{Y_{ij}(t) -\hat{Y}_{ij(-\boldsymbol{\zeta_{ij}})}(t))^T}\boldsymbol{\Psi}\boldsymbol{\zeta_{ij}})] \\
\vspace{2em}
&\propto \mbox{MVN}_L(\boldsymbol{\mu_{\zeta_{ij}}}, \boldsymbol{\Sigma_{\zeta_{ij}}}),
\end{align*}
where
\begin{center}
$\boldsymbol{\mu_{\zeta_{ij}}} = (\boldsymbol{\Psi}^T\boldsymbol{\Psi} + \sigma_{\epsilon}^2(\boldsymbol{\Lambda})^{-1})^{-1}[(Y_{ij}(t) -\hat{Y}_{ij(-\boldsymbol{\zeta_{ij}})}(t))^T\boldsymbol{\Psi}]^T,$ \\
\vspace{1em}
$\boldsymbol{\Sigma_{\zeta_{ij}}} = \sigma_{\epsilon}^2(\boldsymbol{\Psi}^T\boldsymbol{\Psi} + \sigma_{\epsilon}^2(\boldsymbol{\Lambda})^{-1})^{-1},$ 
\end{center}
and 
\begin{align*}
f&(\lambda_l|others) \\
&\propto f(\lambda_l) \prod_{i=1}^{I} \prod_{j=1}^{J} f(\zeta_{ijl}|\lambda_l) \\
&\propto \lambda_l^{-(\alpha_{1l} + \frac{1}{2}IJ + 1)} \times \mbox{exp}(-\frac{\frac{1}{2}\sum_i\sum_j \zeta_{ijl}^2 + \alpha_{2l}}{\lambda_l}) \\
&\propto \mbox{IG}(\frac{1}{2}IJ + \alpha_{1l}, \frac{1}{2}\sum_i\sum_j \zeta_{ijl}^2 + \alpha_{2l}).
\end{align*}

Finally, the conditional posterior distribution for $\sigma_{\epsilon}^2$ is:
\begin{align*}
f&(\sigma_{\epsilon}^2|others) \\
&\propto f(\boldsymbol{y}|others)f(\sigma_{\epsilon}^2) \\
&\propto \mbox{exp}(-\frac{\sum_{i}\sum_{j}\ta{(Y_{ij}(t) -\hat{Y}_{ij}(t))^T(Y_{ij}(t) -\hat{Y}_{ij}(t))}}{2\sigma_{\epsilon}^2}) \times (\sigma_{\epsilon}^2)^{-(\frac{1}{2}~ \ta{IJK} + 1)} \\
&\propto \mbox{Scale-inv-}\chi^2 (\ta{IJK}, \frac{\ta{\sum_{i}\sum_{j}(Y_{ij}(t) -\hat{Y}_{ij}(t))^T(Y_{ij}(t) -\hat{Y}_{ij}(t))}}{\ta{IJK}}),
\end{align*}
where $\mbox{Scale-inv-}\chi^2(\ta{\nu_1, \nu_2})$ is an scaled inverse chi-squared distribution with degrees of freedom $\nu_1$ and scale parameter $\nu_2$.

\bibliographystyle{rss}
\bibliography{Manuscript_HDFMM_CanadianStat}

\end{document}